\documentstyle[prd,aps,floats,epsfig]{revtex}
\newcommand{\be}{\begin{eqnarray}}
\newcommand{\ee}{\end{eqnarray}}
\newcommand{\real}{\mbox{{\rm I\hspace{-2truemm} R}}}
\newcommand{\bol}[1]{\mbox{\boldmath $#1$}}

\newcommand{\ps}{\mbox{$\not\hspace{-0.8truemm}p$}}
\newcommand{\Ds}{\mbox{$\not\hspace{-1truemm}D$}}
\newcommand{\As}{\mbox{$\not\hspace{-1truemm}A$}}
\newcommand{\Bs}{\mbox{$\not\hspace{-1truemm}B$}}
\newcommand{\ad}{\mbox{$\hat a^\dagger$}}
\renewcommand{\a}{\mbox{$\hat a$}}
\begin{document}
\draft
\twocolumn[\hsize\textwidth\columnwidth\hsize\csname
@twocolumnfalse\endcsname
\title{Electromagnetic Contributions to Lepton $g-2$ in a Thick
Brane-World}
\author{R. Casadio$^{a}$, A. Gruppuso$^{b}$ and G. Venturi$^{c}$}
\address{~}
\address{Dipartimento di Fisica, Universit\`a di Bologna,
and I.N.F.N., Sezione di Bologna,
via Irnerio 46, I-40126 Bologna, Italy}
\date{\today}
\maketitle
\begin{abstract}
We consider Standard Model fields living inside a thick four
dimensional flat brane embedded in a (possibly warped) five
dimensional space-time and estimate the electromagnetic
corrections to the anomalous magnetic moment ($g-2$) of the electron
and muon by including virtual massive fermion and gauge boson states.
Constraints on the mass of the ``excited'' states (or thickness of
the brane-world) are obtained.
\end{abstract}
]
\section{Introduction}
Recently there has been a revived interest in models containing
extra spatial dimensions.
The first time this idea was put in concrete form possibly dates
back to the '20s \cite{kaluza}, when the existence of more than
four dimensions was employed in an attempt to unify gravity with
the electromagnetic (EM) field.
In such an approach an electrically charged particle is extended
in the fifth dimension which, because of the relative strength
of gravitational and EM forces, is extremely small
and as a consequence the charged particle is extremely massive
(of order the Planck mass $M_p$).
\par
Higher dimensional spaces also naturally come into play when
strings are considered \cite{green} and one must then compactify
the fundamental theory down to our four dimensional world.
In Refs.~\cite{horava} a compactification scheme (whose low
energy limit is eleven dimensional supergravity) was constructed
for the strongly coupled $E_8\times E_8$ heterotic string.
Six dimensions are then compactified on a Calabi-Yau manifold
and integrated out leaving a five dimensional (bulk)
space-time bounded by two copies of the same D3-brane
\cite{polchinski}.
Matter particles are low energy excitations of open strings
with end-points confined on the D3-brane, which would thus
represent our (brane)world.
Gravitons instead are closed strings and can propagate also in
the extra dimension, which has topology $S/$Z$_2$.
This construction yields a relation between the Planck mass
on the D3-brane and the fundamental string mass scale which
allows the latter to be much smaller than the former,
hence suggesting a solution to the hierarchy problem.
\par
Inspired by this result, several models have been proposed
with various numbers of extra dimensions, which can be
either compact \cite{arkani} or infinitely extended with a
warp factor \cite{randall}.
In both cases, there are parameters which can be tuned to make
the fundamental mass scale small enough to lead to new physics
slightly above 1~TeV without violating Newton's law at the
present level of confidence \cite{long}.
One of the main concerns in such models is to provide a (field
theoretical model) confining mechanism for the matter fields
which does not violate any of the tested properties of the
Standard Model (SM) and yields, at the same time, predictable
effects which can be probed by the forthcoming generation of
detectors \cite{arkani,giudice}.
Early proposals for confining matter fields on a wall of
codimension one are actually older and make use of the
non-vanishing expectation value of a scalar field \cite{rubakov}.
The fact that heavy (``excited'') particle states living in the extra
dimensions have not been detected yet is then generally a
consequence of the small coupling between SM particles and bulk
gravitons, namely ${\mathcal{O}}(1/M_p)$.
\par
In the present letter, we shall consider a model in which the
brane-world has finite thickness (of size $2L_f$ for fermions
and $2L_b\ge 2L_f$ for bosons \cite{schmaltz} which can be
different, reflecting differing confining mechanisms).
The five space-time coordinates are denoted by $\bol{x}$ (or
Greek indices running from $0$ to $3$) for the usual space-time
and $y$ for the extra dimension.
The metric inside the brane is flat Minkowski,
$\eta_{\mu\nu}={\rm diag}\,[-1,+1,+1,+1]$, and matches with an
external (possibly warped \cite{lykken}) space-time metric:
\be
ds^2=a^2(y)\,\eta_{\mu\nu}\,dx^\mu\,dx^\nu+dy^2
\ ,
\ee
where $a(y)=1$ for $|y|<L_b$.
The reason we allow for a warp factor is that we want to consider
just one extra dimension.
Further, the fields we shall study have support inside the (thick)
brane where $a=1$ and possible effects originating in the bulk are
neglected here.
For such a model, we estimate the (order of magnitude of)
EM corrections to the anomalous magnetic moment
of the SM electron and muon and, by comparing with the precision of
the present measurements, obtain bounds for the mass of ``excited''
states which, of course, is related to the thickness of the brane.
\section{Effective four-dimensional action}
Fermion fields $\Psi=\Psi(\bol{x},y)$ couple to the topological
(kink) vacuum of a scalar field $\Phi$ \cite{rubakov}, which we
approximate as
\be
\Phi=\left\{\begin{array}{ll}
-(m_f^2/2)\,L_f &\ \ \ \ \ \ \ y<-L_f
\\
(m_f^2/2)\,y &\ \ \ \ \ \ \ |y|<L_f
\\
+(m_f^2/2)\,L_f &\ \ \ \ \ \ \ y>+L_f
\ .
\end{array}\right.
\ee
Therefore, as we review below, fermions have a confined massless
(chiral) mode, together with a tower of states which are allowed
only if their mass is smaller than $m_f^2\,L_f/2\equiv M_f$
\footnote{
We shall not consider states whose energy exceeds the threshold of
confinement.}.
\par
On neglecting bulk contributions, the five dimensional action for
fermions minimally coupled to gauge bosons in the brane is given by
\be
S_{(5)}=\int_{-L_b}^{+L_b} a^4\,dy\,\int d^4x\,
\bar\Psi\,\left(i\,\Ds-\gamma^5\,\partial_y-\Phi\right)\,\Psi
\ ,
\ee
where $i\,\Ds=-\ps+e\,\As$, $\ps=-i\,\hbar\,\gamma^\mu\,\partial_\mu$
and $e$ is the gauge coupling constant.
Since $\gamma^5=\Pi_L-\Pi_R$ (the difference between left and right
chiral projectors), one can introduce ``creation and annihilation''
operators \cite{schmaltz}
\be
\ad=-{1\over m_f}\,\left(\partial_y-\Phi\right)
\ , \ \ \ \ \
\a={1\over m_f}\,\left(\partial_y+\Phi\right)
\ ,
\ee
such that $[\a,\ad]=1$ and the Lagrangian density becomes
\be
L_{(5)}=
\bar\Psi\,\left(i\,\Ds-m_f\,\a\,\Pi_L-m_f\,\ad\,\Pi_R\right)\,\Psi
\ .
\ee
This allows an expansion for the fermions
\be
\Psi(\bol{x},y)&=&H_0(y)\,\Pi_L\,\psi^{(0)}(\bol{x})
\nonumber \\
&&+\sum_{n=1}^{N_f}
\left[H_{n}(y)\,\Pi_L+H_{n-1}(y)\,\Pi_R\right]\psi^{(n)}(\bol{x})
\ ,
\label{f_exp}
\ee
where $H_{n}$ are the normalized eigenfunctions of the harmonic
oscillator.
Since the zero mode is massless,
\be
(\ps+m_f\,\a)\,H_0\,\Pi_L\,\psi^{(0)}
=H_0\,\ps\,\Pi_L\,\psi^{(0)}=0
\ ,
\ee
$\psi^{(0)}$ can be taken as a two-component Weyl spinor,
$\psi^{(0)}=\Pi_L\,\psi^{(0)}$.
We note in passing that $\a\,H_0=0$ is precisely the equation
which ensures the confinement of the left zero mode within
a width $\ell_f\sim 1/m_f$ around $y=0$.
Since for the (would-be) right zero mode the corresponding
equation $\ad\,\bar R_0=0$ does not admit any (non-vanishing)
normalizable solution in $y\in\real$, we have set
$\Pi_R\,\psi^{(0)}=0$.
\par
\begin{figure}
\centerline{\epsfxsize=3.0in
\epsfbox{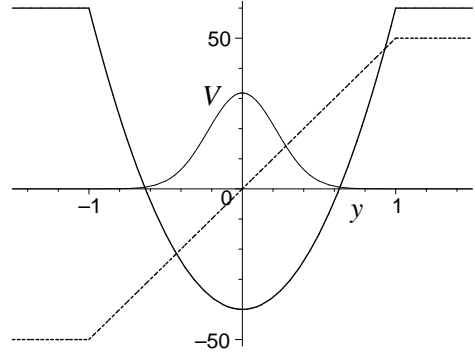}}
\caption{Sketch of the scalar field $\Phi$ (magnified by a factor
of 5; dashed line) and the corresponding confining potential in
Eq.~(\ref{KG}) (continuous line) for $m_f^2/2=10$ (in units with
$L=1$).
The Gaussian curve represents the ground state $H_0$ (magnified by
a factor of 10).}
\label{g2plot01}
\end{figure}

The sum in Eq.~(\ref{f_exp}) ends with a maximum integer
$N_f<\infty$.
The reason for such a cut-off can be easily understood if we
set $\As=0$ and write down the Klein-Gordon equation corresponding
to the Dirac equation obtained from $S_{(5)}$,
\be
&&\left(\ps-m_f\,\ad\,\Pi_L-m_f\,\a\,\Pi_R\right)
\left(\ps+m_f\,\a\,\Pi_L+m_f\,\ad\,\Pi_R\right)\Psi
\nonumber \\
&&=-\left(p^2+m_f^2\,\ad\,\a\,\Pi_L+m_f^2\,\a\,\ad\,\Pi_R\right)
\Psi
\nonumber \\
&&=
-\left[p^2+\left(-\partial_y^2+\Phi^2\right)\right]\,
\Psi=0
\ .
\label{KG}
\ee
It is thus clear that only those modes $\psi_n$ whose eigenvalues
$m_f^2\,n<\Phi^2(L)\equiv M_f^2$ can be retained
(see Fig.~\ref{g2plot01}) and, in the following, we shall consider
only the simplest non-trivial case, that is the lowest level
$n=1$.
\par
Since we need both chiralities to recover the correct low
energy phenomenology, one doubles the fermion fields
$\Psi\to (\Psi_1,\Psi_2)$ and pairs the two zero modes into
one four-component Dirac fermion,
$\psi^{(0)}=(\psi^{(0)}_1,C\,\psi^{(0)}_2)$, where
$C$ denotes charge conjugation.
Further, by introducing an interaction term of the form
\be
L_m=m\,\bar\psi^{(0)}\,\psi^{(0)}
\ ,
\ee
the zero mode can be given the bare mass $m$, which arises as the
vacuum expectation value of the Higgs field, and will have as (free)
propagator
\be
0\ \ \put(1,3){\line(1,0){28}}\hspace{1truecm}\ 0
\ \ \ ={i\over \ps+m}
\ .
\ee
The same mass term could be added for the $n\ge 1$ modes but would be
negligible with respect to $n\,m_f$.
Finally, one can assemble the left spinor components of the
two generations of modes with $n=1$ as
$\psi^{(1)}=(\Pi_L\,\psi^{(1)}_1,C\,\Pi_L\,\psi^{(1)}_2)$
\footnote{We omit writing the analogous rearrangement of the right
components because it does not play any role in the process we are
going to study, see Eq.~(\ref{L4}).},
which will be propagated by
\be
1 \ \ \put(1,3){\line(1,0){28}}\hspace{1truecm}\ 1
\ \ \ ={i\over \ps+m_f}
\ .
\ee
\par
Gauge bosons $A_\mu=A_\mu(\bol{x},y)$ \footnote{It is always
possible to set the fifth component $A_y=0$ as a gauge
choice.}
are assumed to be confined by an analogous mechanism to that of
the fermions and,
on neglecting edge effects, are parametrized by a tower of
Kaluza-Klein~(KK)-like states inside a ``box'' of size $2L_b$
\cite{schmaltz} (in any case, we shall just consider the lowest
states which are not expected to depend on the detailed nature
of the confining mechanism).
Thus, one has a massless ground state and massive KK-like modes
which are allowed provided their mass is smaller than the
confining threshold $M_b$ (possibly different from $M_f$)
\be
A_\mu(\bol{x},y)&=&A_\mu^{(0)}(\bol{x})
+\sum_{n=1}^{N_b}\,B_\mu^{(n)}(\bol{x})\,
{\cos(\pi ny/L_b)\over \sqrt{L_b}}
\nonumber \\
&&
+\sum_{n=1}^{N_b}\,A_\mu^{(n)}(\bol{x})\,
{\sin(\pi ny/L_b)\over \sqrt{L_b}}
\ ,
\label{b_exp}
\ee
where both the modes $A_\mu^{(n)}$ and $B_\mu^{(n)}$ have
an effective four dimensional mass $n\,m_b=\pi\,n/L_b\le M_b$.
We shall again just keep the lowest level $n=1$ and
have the usual massless propagator for $A^{(0)}_\mu$ and the
massive propagator for $A^{(1)}_\mu$ and $B^{(1)}_\mu$,
\be
\begin{array}{l}
0 - - - \,0
\ \ \ =\strut\displaystyle{i\,\eta_{\mu\nu}\over p^2}
\\
\\
1_A - - -\,1_A=
\\
=1_B - - -\,1_B
\ \ \ =i\,\strut\displaystyle{\eta_{\mu\nu}+p_\mu\,p_\nu/m_b^2
\over p^2+m_b^2}
\ .
\end{array}
\ee
\par
\begin{figure}
\centerline{\epsfxsize=3.0in\epsfbox{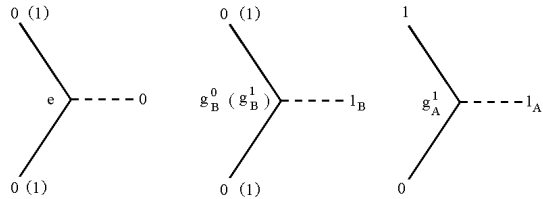}}
\caption{Vertices of the effective theory for $N_f=N_b=1$.}
\label{vertices}
\end{figure}
\begin{figure}
\centerline{\epsfxsize=3.0in\epsfbox{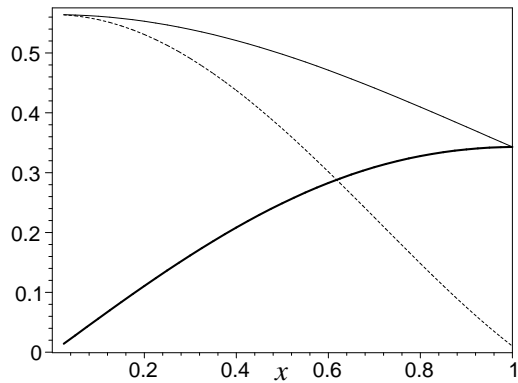}}
\caption{Effective coupling constants (in units of $e$)
as functions of $x\equiv m_b/m_f$:
$g_B^{0}$ (thin solid line); $g_B^{1}$ (dashed line)
and $g_A$ (thick solid line).}
\label{g_eff}
\end{figure}

Upon inserting the expansion (\ref{f_exp}) for the two
pairs of fermions and the gauge field (\ref{b_exp})
into the action $S_{(5)}$, just retaining the lowest levels,
and integrating over the extra dimension, one obtains the
effective four dimensional Lagrangian
\be
L_{(4)}&=&\bar\psi^{(0)}\,\left[\ps+m-e\,\As^{(0)}
-g_B^{0}\,\Bs^{(1)}\right]\,\psi^{(0)}
\nonumber \\
&&+\bar\psi^{(1)}\,\left[\ps+m_f-e\,\As^{(0)}
-g_B^{1}\,\Bs^{(1)}\right]\,\psi^{(1)}
\nonumber \\
&&
-g_A\,\left(\bar\psi^{(1)}\,\As^{(1)}\,\psi^{(0)}
+\bar\psi^{(0)}\,\As^{(1)}\,\psi^{(1)}\right)
,
\label{L4}
\ee
where we have made use of the parity and normalization properties
of the functions $H_n$ in (\ref{f_exp}) and sine and cosine in
(\ref{b_exp}).
We also obtain the five basic interaction vertices displayed in
Fig.~\ref{vertices}, where the effective gauge coupling constants
depend on the overlap of the field modes in the extra dimension,
\be
&&g_B^{0}=e\,\int H_0^2(y)\,{\cos(m_b\,y)\over\sqrt{L_b}}\,dy
\nonumber \\
&&g_B^{1}=e\,\int H_1^2(y)\,{\cos(m_b\,y)\over\sqrt{L_b}}\,dy
\\
&&g_A=e\,\int H_0(y)\,H_1(y)\,{\sin(m_b\,y)\over\sqrt{L_b}}\,dy
\ ,
\nonumber
\ee
and are plotted in Fig.~\ref{g_eff} as functions of the ratio
$m_b/m_f$ which we expect to be related to $L_f/L_b$.
\par
Consideration of additional levels ($n>1$) will lead to analogous
results and even more stringent constraints, that is larger
masses for the allowed ``excited'' states.
However, such contributions are expected to be even more dependent
on the mechanism of confinement.
\section{Anomalous magnetic moment}
We are now ready to compute physical quantities to one loop
order.
Since we are interested in observed particles, the external
legs, besides having on-shell momenta, always correspond to
the observed zero modes $\psi^{(0)}$ and $A^{(0)}_\mu$
(the massive modes $\psi^{(1)}_i$, $A_\mu^{(1)}$ and
$B_\mu^{(1)}$ have never been detected thus far, therefore
$m_f$ and $m_b$ must be at least of order $1\,$TeV
\cite{arkani}).
Further, integrals over internal momenta will be evaluated
with a UV cut-off $\Lambda\sim m_f>m_b$, since the effective
Lagrangian $L_{(4)}$ holds for momenta below the confining
threshold only.
\par
\begin{figure}
\centerline{\epsfxsize=3.0in\epsfbox{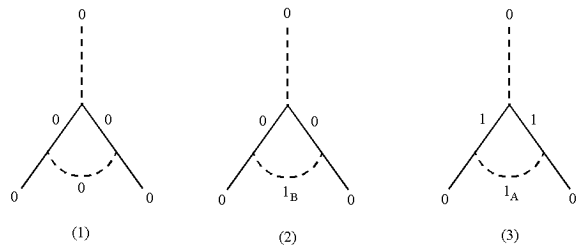}}
\caption{Feynman graphs contributing to $g-2$.}
\label{g-2}
\end{figure}

The EM one-loop contribution to the anomalous magnetic moment
of leptons, $\Delta=(g-2)/2$, is represented by the three
graphs in Fig.~\ref{g-2}.
From all three graphs we extract the form factor corresponding
to the magnetic component in the Gordon decomposition of the
fermion current (see, {\em e.g.}, Ref.~\cite{peskin})
and obtain the following results:
The first graph yields
\be
\Delta_{(1)}={\alpha\over2\,\pi}\,
\left[1+{\mathcal{O}}\left({m^2\over\Lambda^2}\right)\right]
\ ,
\label{d1}
\ee
where the first term is the standard (one-loop) correction and the
second term vanishes when the cut-off $\Lambda\to\infty$;
the second graph gives
\be
\Delta_{(2)}={\alpha\over2\,\pi}\,
\left({g_B^0\over e}\right)^2\,{2\,m^2\over 3\,m_b^2}\,
\left[1+{\mathcal{O}}\left({m^2\over\Lambda^2}\right)\right]
\ ;
\label{d2}
\ee
finally, from the third graph we obtain
\be
\Delta_{(3)}=
{\alpha\over2\,\pi}\,\left({g_A\over e}\right)^2\,
{5\,m\over 12\,m_f}\,\left[1
+{\mathcal{O}}\left({\Lambda-m_b\over \Lambda}\right)
\right]
\ ,
\label{d3}
\ee
which is convergent.
We recall here that $\Lambda\sim m_f$ and notice that
$\Delta_{(3)}$ becomes negligible for $m_b\ll m_f$
because of the dependence on the ratio $m_b/m_f$ of the
coupling constant $g_A$ (see Fig.~\ref{g_eff}).
\par
The measured anomalous magnetic moments of the electron and
muon are in agreement with the SM predictions to a very high level
of precision.
Therefore, the second term in $\Delta_{(1)}$, which arises because
of the cut-off $\Lambda\sim m_f<\infty$, and $\Delta_{(2)}$ and
$\Delta_{(3)}$, which involve virtual $\psi^{(1)}$, $A^{(1)}_\mu$
and $B^{(1)}_\mu$, must be smaller than the experimental error in
$(g-2)/2$ and this implies limits on the possible values of $m_f$ and
$m_b$ or, equivalently, on the thickness of the brane-world.
In particular, $(g-2)/2$ of the electron is measured
with an error $\Delta_e\sim 4\cdot 10^{-12}$ \cite{schwi,rpp}.
Thus, from Eq.~(\ref{d1}) with $m=m_e\sim 0.5\,$MeV and
$\Lambda\sim m_f$ one has that $m_f> 10\,$GeV, which
is however less restrictive than the constraint $m_f,m_b> 1\,$TeV.
The latter constraint also renders the contribution of the second
graph practically negligible, since the coupling constant
$1/3<g_B^0/e<2/3$ and $\Delta_{(2)}$ is thus of the same
order as the second term in $\Delta_{(1)}$.
A stronger prediction instead comes from Eq.~(\ref{d3})
if $m_b\sim m_f$ ($g_A/e\sim 1/3$), namely
$m_f>1.3\cdot 10^7\,m_e\sim 7\,$TeV.
For the muon ($m=m_\mu\sim 100\,$MeV), $(g-2)/2$ is measured with an
error $\Delta_\mu\sim 10^{-9}$ \cite{rpp}.
Eq.~(\ref{d1}) then implies $m_f>100\,$GeV and, from Eq.~(\ref{d3}),
$m_f>5\,$TeV.
Future experiments are expected to lower the error down to
$\Delta_\mu=4\cdot 10^{-10}$ \cite{carey}, which would imply a
limit $m_f>10\,$TeV.
\par
The existence of $\psi^{(1)}$ and massive gauge bosons also gives
new radiative corrections to the lepton ($\psi^{(0)}$) self-mass.
Such terms, because of the longitudinal vector field contribution,
have a leading divergence of the form
$(\Lambda/m_b)^2\,\ln(\Lambda/m_f)$,
which, in contrast to the usual QED case, is multiplied by a factor
proportional to $m_f$ ($\gg m_e$, $m_\mu$).
Clearly, this implies that such contributions could be large.
However, by suitably adjusting $\Lambda$, $m_b$ and $m_f$, one can
keep the corrections finite and small, reabsorbing them in the
definition of the physical mass and the renormalization of external
legs.
This allows one to have a relatively large contribution to $g-2$
without affecting the mass of the light leptons.
\par
Lastly, one may worry about weak corrections: again, in this case,
the masses of the ``excited'' ($W$'s and $Z$) bosons and fermions
will appear [see, {\em e.g.}, our Eq.~(\ref{d3})] and the contributions
will be comparable to our purely EM corrections, thus leading
(barring improbable cancellations) to analogous results.
\section{Conclusions}
In this letter we have considered a model for a thick brane-world
of codimension one and included in the EM one-loop computation of
the anomalous magnetic moment of leptons the contribution of virtual
massive states living inside the brane.
This allows us to put constraints on the possible mass of ``excited''
states in the form of lower limits of the order of $10\,$TeV.
Several simplifying assumptions have been made, in particular we
have just considered one massive mode both for leptons and the
photon.
The results we have obtained could be generalized to include more
massive states (given a detailed model for the confinement) and the
approach we have followed applied to other effects.
\par
We finally observe that, given our construction of a thick brane-world,
our results can be viewed as complementary to those derived from the
inclusion of bulk gravitons \cite{graesser}, SM fields living in the
bulk \cite{rizzo} or other extensions beyond the SM \cite{marciano}.
\acknowledgments
\par
We thank Lorenzo Sorbo and Roberto Zucchini for useful discussions.

\begin{references}
%
%
\bibitem[a]{email}E-mail: casadio@bo.infn.it
\bibitem[b]{email}E-mail: gruppuso@bo.infn.it
\bibitem[c]{email}E-mail: armitage@bo.infn.it
%
%
\bibitem{kaluza}
Th. Kaluza, Sitz. Preuss. Akad., 966 (1921);
O. Klein, Z. Phys. {\bf 37}, 895 (1926).
%
\bibitem{green}
M.B. Green, J.H. Schwarz and E. Witten, {\em Superstring theory}
(Cambridge Univ. Press, Cambridge, England, 1987).
%
\bibitem{horava}
P. Horava and E. Witten, Nucl. Phys. {\bf B460}, 506 (1996);
Nucl. Phys. {\bf B475}, 94 (1996); E. Witten, Nucl. Phys. {\bf B471},
135 (1996).
%
\bibitem{polchinski}
J. Polchinski, hep-th/9702136.
%
\bibitem{arkani}
N. Arkani-Hamed, S. Dimopoulos and G. Dvali, Phys. Lett. B {\bf 429},
263 (1998); Phys. Rev. D {\bf 59}, 0860004 (1999);
I. Antoniadis, N. Arkani-Hamed, S. Dimopoulos and G. Dvali,
Phys. Lett. {\bf B436} (1998) 257.
%
\bibitem{randall}
L. Randall and R. Sundrum, Phys. Rev. Lett. {\bf 83}, 3370 (1999);
{\bf 83}, 4690 (1999).
%
\bibitem{long}
J.C. Long, H.W. Chan and J.C. Price, Nucl. Phys. {\bf B539}, 23 (1999).
%
\bibitem{giudice}
G.F. Giudice, R. Rattazzi and J.W. Wells, Nucl. Phys. {\bf B544}, 3
(1999).
%
\bibitem{rubakov}
V.A. Rubakov and M. Shaposnikov, Phys. Lett. {\bf B 125}, 136 (1983);
M. Visser,  Phys. Lett. {\bf B 159}, 22 (1985).
%
\bibitem{schmaltz}
N. Arkani-Hamed and M. Schmaltz, Phys. Rev. D {\bf 61} 033005 (2000).
%
\bibitem{lykken}
J. Lykken, R.C. Myers and J. Wang, hep-th/0006191.
%
\bibitem{peskin}
M.E. Peskin and D.V. Schroeder, {\em An introduction to quantum
field theory}, (Perseus Books, Cambridge, Massachusetts, 1995).
%
\bibitem{schwi}
P. Schwinberger, R. Van Dyck Jr. and H. Dehmelt, Phys. Rev. Lett.
{\bf 59}, 26 (1987).
%
\bibitem{rpp}
Particle data group, {\em Review of particle physics}, Eur. Phys.
J. {\bf C3}, 1 (1998).
%
\bibitem{carey}
R.M. Carey et al., Phys. Rev. Lett. {\bf 82}, 1632 (1999).
%
\bibitem{graesser}
M.L. Graesser, Phys. Rev. D {\bf 61}, 074019 (2000).
%
\bibitem{rizzo}
H. Davoudiasl, J.L. Hewett and T.G. Rizzo, hep-ph/0006097.
%
\bibitem{marciano}
W.J. Marciano, hep-ph/9902332.
%
\end{references}
\end{document}